\documentstyle[12pt,epsfig]{article}

\newcommand{\dr}{\displaystyle \partial R}
\newcommand{\dh}{\displaystyle \partial H}
\newcommand{\hs}{\displaystyle \bf {H}}
\newcommand{\as}{\displaystyle \bf {a}}
\newcommand{\bs}{\displaystyle \bf {b}}
\newcommand{\cs}{\displaystyle \bf {c}}
\newcommand{\hn}{\displaystyle H_1^{(n)}}
\newcommand{\ho}{\displaystyle H_1^{(1)}}
\begin{document}

\large

\begin{center}
\Large \bf
           SQUID nature of microwave absorption in a high-$T_c$ \\
           superconducting Ho-Ba-Cu-0 single crystal
\end{center}

\vspace{2mm}

\begin{center}
\bf M.K.Aliev, G.R.Alimov \footnote[1]{Corresponding author.
e-mail: gleb@iaph.silk.org }, I.Kholbaev, T.M.Muminov,
and B.A.Olimov
\end{center}

\begin{center}
\small Institute of Applied Physics, Tashkent State University
700095, Tashkent, Uzbekistan \\
\end{center}

\begin{abstract}
A high-$T_c$, superconducting $Ho_1 Ba_2 Cu_3 O_{7-x} $ single crystal with
$T_c = 86.8 K$ is investigated on a modified ESR spectrometer at temperatures
$T>78 K$. A signal is observed, whose spectrum in the range of magnetic
fields $H<0.7 Oe$ has the form of equivalent absorption lines with an
interval $\Delta = 9 \times 10^{-3} Oe$. Detailed measurements of the
dependence of the spectrum on the amplitude of the microwave field, the
temperature, and the orientation of the single crystal are
reported. It is established that the absorption spectrum exhibits a typical
behavior inherent in SQUIDs with a critical current that decays linearly
with increasing T. Three different techniques are used to determine the
so-called hysteresis parameter of the postulated SQUID: $Li_c /\Phi_0 $
(L and $i_c$ are the inductance and critical current of the SQUID, and
$\Phi_0$ is the quantum of magnetic flux). All three techniques give the
same result: $Li_c (T=79 K)/\Phi_0= 3.1\pm 0.3$, attesting to the
existence of a SQUID-like structure in the single crystal.
\end{abstract}

\vspace{1mm}

\section{Introduction}
\indent
Since the advent of high-temperature (high-$T_c$)
superconductivity, the ESR spectrometer has found application for
the investigation of microwave absorption in high-$T_c$ superconductors
in addition to its conventional use. One of the
most interesting phenomena discovered by means of ESR
spectrometers is a periodic dependence of the microwave
absorption in high-$T_c$ single crystals of the 1-2-3 type on
the external magnetic field H (the period
$\Delta H\approx 10^{-2}Oe$) \cite{ref1} - \cite{ref8}.
Despite consensus that this phenomenon is a macroscopic
quantum effect, a single, unambiguous explanation has yet to
be found for the microwave absorption mechanism itself. It
has been hypothesized \cite{ref4} - \cite{ref6} that the phenomenon
is attributable to the existence of structures resembling SQUIDs in
high-$T_c$ superconductors. We now propose to confirm the validity
of this hypothesis on the basis of quantitative self consistency,
within the SQUID model, of the sum-total of
experimental data obtained from more detailed measurements.
\section{Experimental Procedure}
\indent

A RADIOPAN SE/X-2543 ESR spectrometer ($\nu\approx 9 GHz,
 P_{max} = 130 mW $) with a $TE_{102}$ cavity was used in the
experiment. A magnetic field was generated by Helmholtz
coils. The cavity was positioned away from the electromagnet
of the spectrometer and together with the Helmholtz coils
was placed in a magnetostatic shield to achieve a hundredfold
suppression of the earth's magnetic field.

We designed a thermal regulation system capable of
holding a sample at a steady temperature anywhere in the
range $78 K< T< 100 K$ within $\pm 0.1 K$ error limits. The
conventional field geometry for ESR spectrometers was used in
the experiment: $ \hs\perp\hs_1 $ ($\hs$ is the static field, and
$\hs_1$ is the microwave field).

The investigated sample was a $Ho_1 Ba_2 Cu_3 O_{7-x} $ single
crystal in the form of a $ 1.0\times0.7\times0.1mm^3 $ rectangular wafer.
The plane of the waver coincided with the crystallographic
($\as\bs $) plane, and the crystallographic $ \cs $ axis was parallel
to the shortest edge of the single crystal. The measurements were
performed with the single crystal oriented so that $\cs\perp\hs_1$, and
the angle $\varphi$ between $\cs$ and $\hs$ could be varied by rotating the
single crystal about the direction of $\hs_1$.

The microwave absorption signal was recorded by a
modulation technique with the H field modulated at the frequency 
$\nu_m = 100 kHz$. From now on we denote this signal by
"$\dr /\dh$" (R is the absorption), using quotation marks to
underscore the conditional nature of this notation: In view of
the steepness of the dependence R(H), in general the H-field
modulation amplitudes used by us ($h_m \geq 1.25 mOe $) were not
small enough for the recorded signal to be interpreted as the
derivative of the absorption in the strict sense.

Prior to each consecutive measurement of the signal
spectrum the sample was heated above the critical transition
temperature and then cooled to necessary temperature in zero field
($H\sim 1 mOe$).

The critical temperature of the investigated sample, determined 
from the position of the superconducting transition
peak observed in the temperature dependence of the signal in
fields $H>30 Oe$ with high modulations of the H field, had a
value $T_c =86.8\pm 0.2 K$ (for more on the superconducting
transition peak see Refs. \cite{ref9,ref10} ).

\section{Experimental Results}
\indent

For low field modulations $h_m =10^{-3} - 10^{-2} Oe $ and microwave 
power inputs to the cavity exceeding the threshold value determined 
for a given temperature, the observed signal oscillates as the field H 
is varied. The period of the signal oscillations with respect to H does 
not depend on the microwave power and changes only when the crystal is 
rotated. The minimum oscillation period is attained at the angle
$\varphi = 90^{\circ}$, when it has the value $\Delta H = 9 \times 1O^{-3} Oe$.
The field dependence of the signal becomes irregular in the vicinity of
$H\approx 0.7 Oe$, where a comparison of the spectra for forward
and backward scans shows that hysteresis effects, having
been negligible near H = 0, intensify as the retrace point in
scanning approaches the indicated value of the field.
\begin{figure}
\vskip2.5cm
\begin{center}
\hspace*{-2.cm}
\parbox{7cm}{\epsfxsize=6.cm \epsfysize=7.cm \epsfbox [5 5 500 500]
{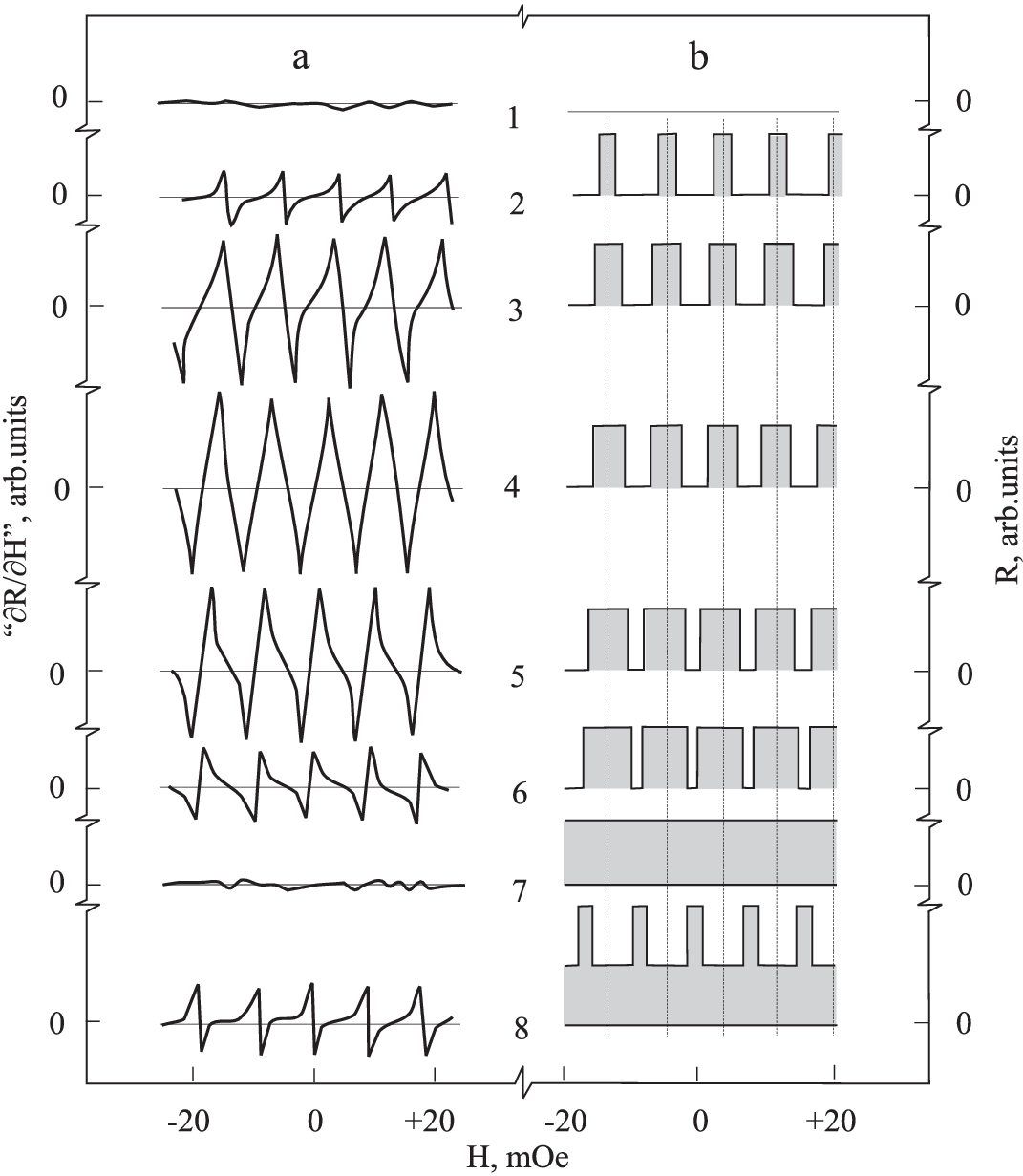}{}}
\parbox{7cm}{\epsfxsize=7.cm \epsfysize=6.cm \epsfbox [5 5 500 500]
{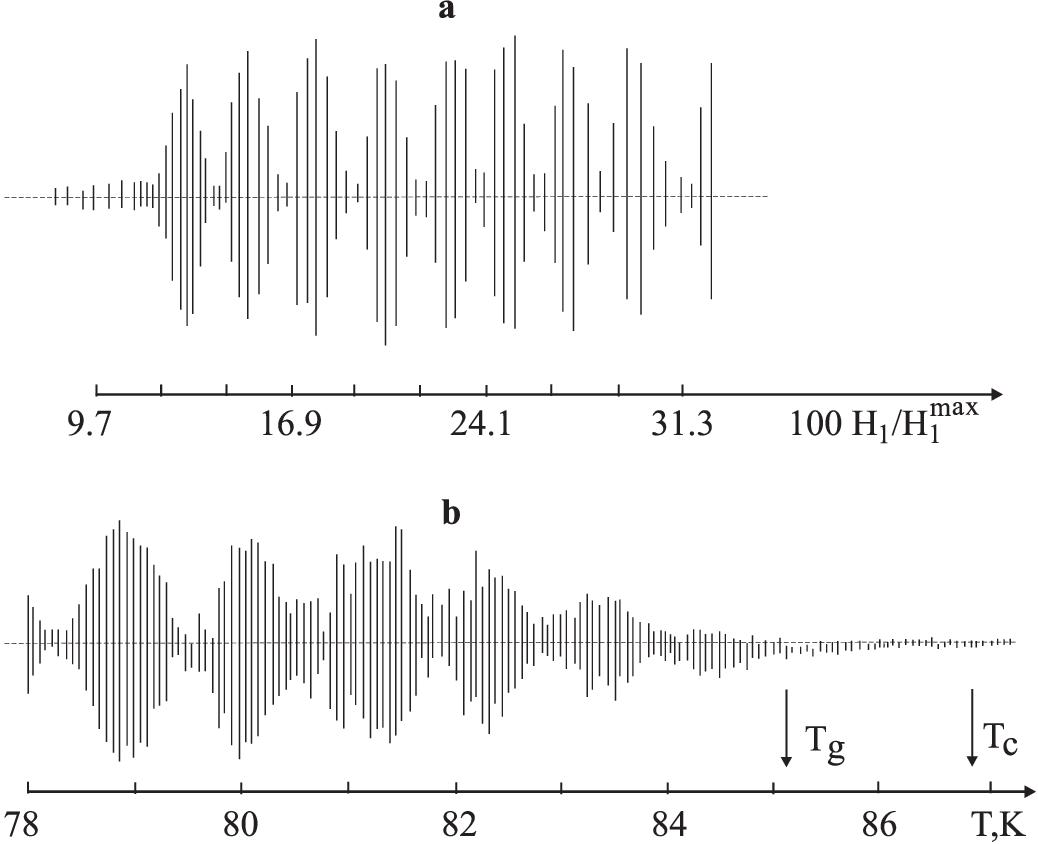}{}}
\end{center}
\begin{center}
\vspace*{-1.5cm}
\hspace*{-1.cm}
\parbox{7.5cm}{\small
Fig.1. Evolution of the absorption spectrum as the microwave power is
increased at T= 79 K. a) Variation of the spectrum of the "$\dr /\dh $"
signal; b) corresponding absorption spectrum R(H) in the SQUID model.
The spectra are numbered in sequence as the microwave power is uniformly
increased from -18.4 dB to - l6.5 dB (0 dB corresponds to $ P_{max}$ = 130
mW).}
\hspace*{.5cm}
\vspace {-0.5cm}
\parbox{7.5cm}{\small
Fig.2. Oscillation amplitude of "$\dr /\dh $" signal vs microwave field at
T= 79K (a) and vs temperature at P = - 14.8 d8 (b). $H_1^{max}$ is the
amplitude of the microwave field at $P_{max}$= 130 mW (0 db), the dashed
lines correspond to zero signal level, and the amplitudes are given in
arbitrary units. \\
\phantom{gggg} \\
\phantom{gggg} }
\end{center}
\end{figure}

Unlike the oscillation period, the amplitude and the H
dependence of the form change abruptly as the microwave
power is varied. In Fig. Ia this behavior is reflected by the
display of several graphs of the field dependence of the signal 
near H=O at T=79 K and $\varphi =90^{\circ}$, plotted for various
values of the microwave power. It is evident from the figure
that, as the microwave power is increased, the amplitude of
the signal oscillations increases from zero (curve 1a), passes
through a maximum (curve 4a), and then decays to zero
(curve 7a). With a further increase in the microwave power
the amplitude once again increases (curve 8a) and, as shown
by subsequent measurements, the indicated dependence repeats 
periodically, showing up as one new series after another. 
We have traced 15 such series in detail.

We now call attention to the change in the form of the H
depcndence of the signal. It is evident from Fig. 1a that one
of the distinguishing features of this dependence is found in
the "shoulders," i.e., the sloping parts of the curves connecting 
the maxima and minima of the signal. Whereas the
shoulders of curves 2a and 3a occur to the left of the
maxima, those of curves 5a and 6a have already shifted to
the right of them. In this sense curve 4a represents a transition 
stage in that it has a symmetric profile. In this sense curve 4a
represents a transition stage in that it has a symmetric profile.
The shoulders of curve 8a, which launches a new series of spectra, 
are once again to the left of the maxima. A careful examination of the
curves reveals that this change in their, profile is associated
with movement of the maxima and minima as the microwave
power is increased, the maxima shifting to the left, and the
minima to the right. This movement of the extrema is clearly
evident in curves 2b-6b, where the vertical bars indicate the
positions of the maxima and minima of the signal, which
correspond to the vertical bars forming the left and right
boundaries, respectively, of the hatched part of the curves.
Using these curves, we can also readily conclude that the
intervals between the maxima and minima of the signal increase 
uniformly from zero to the oscillation period $\Delta H$
within a single series (curves 1-7). And one final, important
detail: Curve 8a(b), which is the start of a new series of
spectra, is shifted a half-period, i.e., $\Delta H /2$, relative to the
corresponding curve 2a(b) in the first series. We have observed this shift 
in transition from one series to the next in all 15 of the measured series.

Figure 2a shows the dependence of the amplitude of the
signal oscillations on the amplitude of the microwave field
$H_1 $ ($ H_1 \sim P^{1/2} $, where P is the microwave power input).
The ends of each of the vertical lines drawn in this figure signify
the maximum and minimum values of the signal as a function of H at 
the corresponding value of $H_1$. The formation of
the above-discussed series is clearly evident from the figure.
Of special note is the interesting fact that the threshold values 
of $H_1$ with which the series begin and which we denote
from now on by $\hn $ (n = 1, 2,... enumerates the series)
are separated by equal intervals $\Delta H_1$ and can be expressed
by the arithmetic progression
\begin{equation} \label{f1}
  \hn=\ho+(n-1)\Delta H_1,
\end{equation}
where $H_1 $ is the threshold of the first series and is simultaneously 
the absolute signal-generation threshold. It is also evident
from the figure that the maximum signal of the series
does not show any overall tendency to increase or decrease
with increasing sequence number n, but merely oscillates in
the nature of large-scale modulations with respect to $H_1 $.
Since the measurements were performed over a wide range
of microwave power, we found it necessary to work with
both square-law and linear detection regimes. The first nine
series shown in Fig. 2a have been obtained in an almost
square-law regime, where the dependence of the recorded
signal on the microwave power P is known to correspond to
the true P dependence for 
$ \Delta R $ $ (\Delta R\sim R(H+h_m)-R(H-h_m)) $. The next six series
were measured directly in a near-linear regime. Here the maximum signal 
is observed to decay from one series to the next. Not to be dismissed is
the possibility of attributing this decay to the instrumental distortion 
factor $ ^{-1/2} $, which is well known in ESR spectroscopy and is inherent
in the P dependence of the signal when linear detection is used. 

The investigation of the temperature dependence of the signal spectrum in 
the interval $78 K< T< T_c$ gives the following results. The oscillation 
period $\Delta H$ remains constant as the temperature is varied in this 
interval. The thresholds $ \hn $ shift uniformly toward lower (higher) 
values of $ H_1 $ as the temperature is raised (lowered), so that the 
interval between thresholds $ \Delta H_1 $ also remains constant. 
As T increases, the signal spectrum for a given $ H_1 $ goes through 
the same evolutionary phases with the formation of series observed as
$ H_1 $ increases at a fixed temperature T (Fig. 1). The formation
of temperature series is distinctly perceptible in Fig. 2b,
which shows the temperature dependence of the signal oscillation amplitude 
at a fixed $ H_1 $ (the vertical lines in the figure
have the same significance as in Fig. 2a except that now they
correspond to different temperatures). To distinguish between the two types 
of series from now on, we designate them as a $ H_1 $ series (series of 
spectra ordered on $ H_1 $ ) and a T series (series of spectra ordered on T).

It is evident from a comparison of Figs. 2a and 2b that the main qualitative
difference between the T series and the $ H_1 $ series is that the maximum 
signal of the T series decreases monotonically toward higher temperatures 
with increasing sequence number. It is noteworthy that the temperature at 
which the signal vanishes, $T_g = 85.2 K$, is consiclerably lower than the 
critical temperature $T_c = 86.8 K$. Also remarkable is the approximate 
equality of the intervals between thresholds of the T series with 
$\Delta = 1.1-1.3 K$. It has been established that the temperature thresholds
shift uniformly toward lower (higher) temperatures as $ H_1$ increases 
(decreases), where a negative (positive) shift in the case of sufficiently 
large variations of $ H_1$ is accompanied by the emergence (disappearance) 
of new (existing) series at a constant temperature $T_g$. In the case of 
sufficiently low values of $H_1$ we observe an absolute temperature threshold
for the appearance of a signal; as $H_1$ increases, this threshold
shifts toward lower temperatures until it attains liquid nitrogen level, 
after which it is not longer observable. This is the case depicted in 
Fig. 2b, where the first observable low-temperature threshold is the 
threshold of the third T series. We conclude, therefore, that the threshold
of the T series can be expressed, as in the case of the $H_1$ series, by
an arithmetic progression.
\begin{equation} \label{f2}
  T^{(n)}=T^{(1)}+(n-1)\Delta T,
\end{equation}
where n = 1, 2, 3,..., N (N is the highest sequence number, which depends 
on $H_1$), and $ T^{(1)} $ is simultaneously the threshold of the first T
series and the absolute temperature threshold of signal generation at a 
given $H_1$.

As we mentioned at the outset, the signal oscillation period $ \Delta H $ 
in our experimental geometry depends on the angle $ \varphi $ between 
the crystallographic $\cs$ axis and the direction of the field H. By 
rotating the single crystal about the direction of the microwave field 
$H_1$ and measuring the signal oscillation period $\Delta H$ for various 
angles $ \varphi $ we have established the functional relation
\begin{equation} \label{f3}
  \Delta H(\varphi)=\Delta H(90^{\circ})/\sin\varphi,
\end{equation}
where $ \Delta H (90^{\circ})=9\times 10^{-3} Oe. $ This fact indicates 
that a preferred direction parallel to the crystallographic ($ \as \bs $)
plane exists in the single crystal and that the external field influences
the absorption signal in such a way that only its projection onto this 
direction is effective.

\section{Discussion of the results and conclusions}
\indent

All the experimental results obtained in our work can be
explained on the hypothesis that a superconducting ring with
a weak link, i.e., a SQUID, exists in the single crystal. The
first evidence in support of this hypothesis can be found in
the form of the absorption spectrum R(H) and its evolution
as the microwave power is increased, as discerned-from the
SQUID model and shown in Fig. 1b in accordance with the
measured "$\dr /\dh$" signal spectra. The fact that the
"$\dr /\dh$" signals in the extreme spectra of each $H_1$ series
are vanishingly small can be attributed to the fact that the
modulation amplitude by virtue of its finiteness greatly exceeds the 
width of the "absorption line" of the spectrum R(H) at the beginning 
of the series, and greatly exceeds the width of the "gap" between them 
at the end of the series. The formation of the spectra of the $H_1$ 
series is attributable to the actual mechanism of microwave power 
absorption in the SQUID when it operates in a hysteresis regime. This
mechanism consists in the fact that each entry or exit of a
fluxon in a SQUID ring, stimulated by microwave field oscillations, 
has a jumplike behavior and is accompanied by energy dissipation in 
the weak link. The sequence number n in an $H_1$ series in the SQUID 
model indicates the number of successively entering (departing) fluxons 
involved in the formation of the absorption maxima of the spectra R(H) 
of the given series (in this sense the minima correspond to n-1 fluxons)
within one oscillation period. Remarkably, the increase in the absorption 
from one series to the next does nothing more than raise the homogeneous 
"background" of the spectrum R(H), whereas the intensity of the part of 
the spectrum that varies with H, i.e., the "absorption line," remains 
unchanged. This fact accounts for the observed approximate invariance of 
the maximum of the "$\dr /\dh$" signal with increasing sequence number of 
the series. It is evident from Fig. 1b that the indicated homogeneous 
"background" is formed as a result of the spreading of the "absorption line"
at the end of each series, and it must therefore be cumulative with 
increasing sequence number of the series.

It follows from the SQUID model that the centers of the
"absorption line" in the spectra of consecutive series must
he shifted relative to one another by a half-period, where the
center of one line in series with even n must be situated at
the point H = 0. Both of these conjectures are consistent with
our measurements.

Finally, according to the SQUID model,
\begin{equation} \label{f4}
 \Delta H=\Phi_i /[S \cdot\cos
 \widehat{({\bf\mathstrut H},{\bf\mathstrut n})}],
\end{equation}
($\Phi_0$ is the quantum of magnetic flux, S is the area of the
SQUID, and {\bf n} is the normal to the face of the SQUID); this
result agrees with our observed relation (\ref{f3}), provided only
that we consider {\bf n} to be situated in the crystallographic
($\as \bs$) plane.

We now attempt to verify quantitatively the validity of the SQUID model
of microwave absorption. For the SQUID in this case we invoke the linear
theory of Silver and Zimmerman \cite{ref11}. According to this theory, 
the thresholds of the $H_1$ -series must be described by Eq. (\ref{f1}), 
where the quantities $\Delta H_1$, and $\ho$ in it are expressed as 
follows in terms of the
basic SQUID parameters:
\begin{equation} \label{f5}
 \Delta H_1=\Phi_0/[2S \cdot 
\cos\widehat{({\bf H_1},{\bf n})}],
\end{equation}
\begin{equation} \label{f6}
  H_1^{(1)}(T)=\Delta H_1\cdot [2Li_c(T)-\Phi_0]/\Phi_0,
\end{equation}
where L and $i_c$ are the inductance and critical current of the
SQUID, respectively.

We make the following departure in connection with
Eqs. (\ref{f4}) and (\ref{f5}). Our experimental geometry is such that the
direction of {\bf n} cannot be determined exactly. The experimental data 
expressed by Eq. (\ref{f3}) permit only the statement that
the vector {\bf n} is situated in the ($\as\bs$) plane. However, there is
an indirect technique for amplifying the specification of this
vector. When the single crystal is oriented with
$\varphi\equiv \widehat{({\bf\mathstrut H},
{\bf\mathstrut c}})=90^{\circ} $, the ($\as\bs$) plane is congruent with the
($\widehat{({\bf\mathstrut H_1},{\bf\mathstrut H})}$ ) 
plane, so that $\cos \widehat{({\bf\mathstrut H},{\bf\mathstrut n}})=
\sin \widehat{({\bf\mathstrut H_1},\bf{\mathstrut n}})$.
It then follows from (\ref{f4}) and (\ref{f5}) that
$$\tan \widehat{({\bf\mathstrut H},{\bf\mathstrut n}})=
\Delta H/2 \Delta H_1. $$
It is evident from Figs. 1 and 2a that for the indicated orientation of 
the single crystal $\Delta H=9 \cdot 10^{-3} Oe $  and
$\Delta H_1=2,4 \cdot 10^{-2} \cdot H_1^{max} $. Approximating 
$ \rm  H_1^{max} $  roughly by its calculate value in the case of 
the unloaded cavity, $ H_1^{max} \approx 1,6 Oe $, we find that
$$\tan\widehat{({\bf\mathstrut H},{\bf\mathstrut n}})\approx 10^{-1}$$. 
The smallness of the resulting number implies that the vector
{\bf n} lies in practically the same direction as the vector {\bf H}.
In our experiment, for the given orientation of the single crystal, its
two small faces parallel to the {\bf c} axis were perpendicular to
{\bf H}. It is reasonable to infer, therefore, that the vector {\bf n} is
perpendicular to these faces and, accordingly, the plane of
the SQUID ring is parallel to it. Setting 
$\cos\widehat{({\bf\mathstrut H},{\bf\mathstrut n}})\approx 1 $
in (\ref{f4}), we can also determine the area of the SQUID:
$$ S\approx\frac{\Phi_0}{\Delta H}\approx 2\cdot 10^{-3} mm^2.$$
A comment needs to be interjected at this point to avoid
misunderstanding. In the experiment we have a situation
similar to the case ${\bf H_1}\bot {\bf n}, {\bf H} \| {\bf n} .$ 
If these exact conditions
were met, then observing a SQUID signal would require the
input of infinite microwave power, because Eqs. (\ref{f5}) and (\ref{f6})
give divergent quantities in this case. In reality, however,
there is always a small error in the mounting of the sample in
the holder. The value of $\tan\widehat{({\bf\mathstrut H},
{\bf\mathstrut n}})\approx 10^{-1} $ found by us corresponds to 
$\sim 5^\circ $,  which is fully admissible in the mounting of the 
sample in view of its small dimensions. This is why we have obtained finite
values of $\Delta H_1$, in the experiment, which are nonetheless large 
enough to ensure high accuracy in the determination of this quantity. It 
must be noted that the indicated error does not in any way affect the
conclusions that follow, because the equations used by us are
general.

The existence of the T series of absorption spectra is also
readily explained in the SQUID model. The critical current
of Josephson junctions is known to decrease monotonically
as the temperature increases. Accordingly, it follows from
Eqs. (\ref{f6}) and (\ref{f1}) that the threshold values $\hn$ must shift
uniformly toward lower values with increasing temperature,
so that the absorption spectrum for a fixed $H_1$ goes through
sequential stages of evolution in the same way as then $H_1$ is
increased at a fixed temperature. Obviously, the threshold
values $ T^{(n)} $  for a given $H_1$ are given by the equation
\begin{equation} \label{f7}
  \hn(T^{(n)})=H_1.
\end{equation}

It follows from (\ref{f7}) that equation (\ref{f2}) describing our 
experimentally determined thresholds of the T series is valid in
the SQUID model if the critical current $ i_c $ decreases with increasing
temperature according to the linear law 
\begin{equation} \label{f8}
  i_c(T)=\alpha\cdot(T^* -T)
\end{equation}
($\alpha $ and $ T^* $ are constants), where the quantities $\Delta T $ 
and  $ T^{(1)} $ in (\ref{f2}) are now given by the expressions
\begin{equation} \label{f9}
  \Delta T =\Phi_0 /2L\alpha,
\end{equation}
\begin{equation} \label{f10}
  T^{(1)}(H_1)=T^* -\Delta T\cdot (1+ H_1 / \Delta H_1).
\end{equation}
The amplitude of the signal oscillations is expressed as
follows in the SQUID model:
\begin{equation} \label{f11}
{(" \partial R/\partial H ")}^{max} \sim 2 \nu [2Li_c(T) -\Phi_0]
\cdot \Phi_0/2L,
\end{equation}
from which it follows that the intensity of the signal must decay as the 
temperature increases. The practically linear decay of the signal observed 
in Fig. 2b corresponds to condition (\ref{f8}). The temperature $T_g$ at 
which the signal is observed to vanish has a special significance as the 
temperature for which
\begin{equation} \label{f12}
  Li_c(T_{\rm g})/\Phi_0=1/2
\end{equation}
(At $ T > T_g $ the SQUID already operates in the zero-hysteresis
regime, for which the given absorption mechanism no longer
functions). Equation (\ref{f12}) with (\ref{f8}) and (\ref{f9}) leads to 
the relation
\begin{equation} \label{f13}
  T^*=T_{g}+\Delta T.
\end{equation}
Substituting the experimentally determined values of $ T_{g}=85.2\pm 0,1 K $
and $ \Delta T=1.2\pm0.1 K $, we find that $ T^* =86.4\pm 0.2K $.
This value of $T^* $ practically coincides with the measured critical 
temperature $ T_c =86,8\pm 0,2Š $ consistent with the theory of Josephson 
junctions \cite{ref12}.

Using equations, written above can easily obtain three expressions for the
"hysteresis parameter" $ Li_c (T)/\Phi_0 $  in terms of
various sets of experimentally determinated quantities:
\begin{equation} \label{f14}
  [(N-n)+1]/2\le Li_c (T^{(n)})/\Phi_0<[(N-n)+2]/2,
\end{equation}
where $ T^{(n)} $ denotes the threshold temperatures, and N is the maximum 
sequence number of the T series observed for some specific value of $H_1$;
\begin{equation} \label{f15}
  Li_c (T)/\Phi_0 =1/2[1+(T_{g}-T)/\Delta T];
\end{equation}
\begin{equation} \label{f16}
  Li_c (T)/\Phi_0 =1/2[1+\ho(T)/\Delta H_1].
\end{equation}

We can use Eqs. (\ref{f14}) - (\ref{f16}) to test the self-consistency of
our experimental data. For example, in Fig. 2b the first visible threshold
of the T series corresponds to T= 78.2 K, a simple computation of the total
number of subsequent thresholds yields (N-n)=5, and hence, according to 
(\ref{f14}), we have
\begin{equation} \label{f17}
  Li_c (T=78.2K)/\Phi_0 =3.0 \div 3.5.
\end{equation}
For comparison we give the result of estimating the hysteresis parameter 
obtained at the same temperature from Eq.(\ref{f15}) using the experimental
data for $\Delta T$ and $ T_g $  ($ \Delta T=1.2 \pm 0.1K, $
and $ T_g =85.2 \pm 0.1K $):
\begin{equation} \label{f18}
                  Li_c (T=78.2K)/\Phi_0 =3.5 \pm 0.3.                    
\end{equation}
Clearly, the results of both estimates agree.

Continuing the test for self-consistency of our experimental data, we now 
determine the hysteresis parameter by means of (\ref{f16}), using the 
data from Fig. 2a: T=79 K and 
$ \ho / \Delta H_1 =(12.1 \pm 0.3)/(2.4 \pm 0.1). $
As a result of simple calculations we obtain
\begin{equation} \label{f19}
                  Li_c (T=79K)/\Phi_0 =3.0 \pm 0.2.                      
\end{equation}
On the other hand, it is readily verified that Eq. (\ref{f15}) gives at
T=79 K
\begin{equation} \label{f20}
                  Li_c (T=79K)/\Phi_0 =3.1 \pm 0.3.                      
\end{equation}
We see, therefore, that our experimental data are quantitatively 
self-consistent within the framework of the SQUID model. In our opinion,
this fact is the most conclusive evidence in favor of the SQUID nature 
of microwave absorption in type 1-2-3 high-$T_c$ superconducting single 
crystals.

\end{document}